\documentclass[preprint,showpacs,showkeys,aps,prb]{revtex4}
\usepackage[english]{babel}
\usepackage[dvips]{graphicx}

\begin{document}

\title{Steady Periodic Shear Flow {\it is} Stable in Two Space Dimensions .
Nonequilibrium Molecular Dynamics {\it versus} Navier-Stokes-Fourier
Stability Theory -- A Comment on two Arxiv Contributions .} 

\author{
Wm. G. Hoover and Carol G. Hoover               \\
Ruby Valley Research Institute                  \\
Highway Contract 60, Box 601                    \\
Ruby Valley, Nevada 89833                       \\
}

\date{\today}



\vspace{0.1cm}

\begin{abstract}
Dufty, Lee, Lutsko, Montanero, and Santos have carried out stability analyses of
steady stationary shear flows.  Their approach is based on the compressible and
heat conducting Navier-Stokes-Fourier model.  It predicts the unstable exponential
growth of long-wavelength transverse perturbations for both two- and three-dimensional
fluids.  We point out that the patently-stable two-dimensional periodic shear flows
studied earlier by Petravic, Posch, and ourselves contradict these predicted instabilities.
The stable steady-state shear flows are based on {\it nonequilibrium}
molecular dynamics with simple {\it thermostats} maintaining nonequilibrium
stationary states in two space dimensions.  The failure of the stability analyses
remains unexplained.

\end{abstract}

\maketitle

\section{Introduction}

It recently came to our attention that the {\it stability} of high-Reynolds'-number atomistic
shear flows\cite{b1,b2,b3,b4,b5} apparently contradicts the {\it instability} predicted by a
perturbation analysis\cite{b6,b7} of the Navier-Stokes-Fourier equations for steady
compressible heat-conducting simple shear.  Here we develop an alogorithm for solving
the Navier-Stokes-Fourier problem directly.  We also apply a two-dimensional
version of the perturbation analysis to soft disks.  This work confirms the
contradictory nature of the microscopic and macroscopic approaches.  The failure of
the perturbation analysis remains unexplained.

In Section II we describe the atomistic and continuum approaches to simulating simple
shear.  We apply the continuum algorithm\cite{b8,b9} to a Navier-Stokes-Fourier gas-phase
model of simple shear.  In Section III we formulate a dense-fluid constitutive
model\cite{b3,b10,b11} for the {\it continuum} description of the soft disks used in
{\it atomistic} simulations.
In Section IV we apply several versions of the continuum perturbation theory to this
soft-disk constitutive model.  The results of that theory are contradicted by the
existing molecular dynamics results.  Section V sums up the current state of our
knowledge, emphasizing the open problem which remains unsolved.

\section{Steady Shear Flows in Two Space Dimensions}

\begin{figure}
\includegraphics[height=11cm,width=8.5cm,angle= -90]{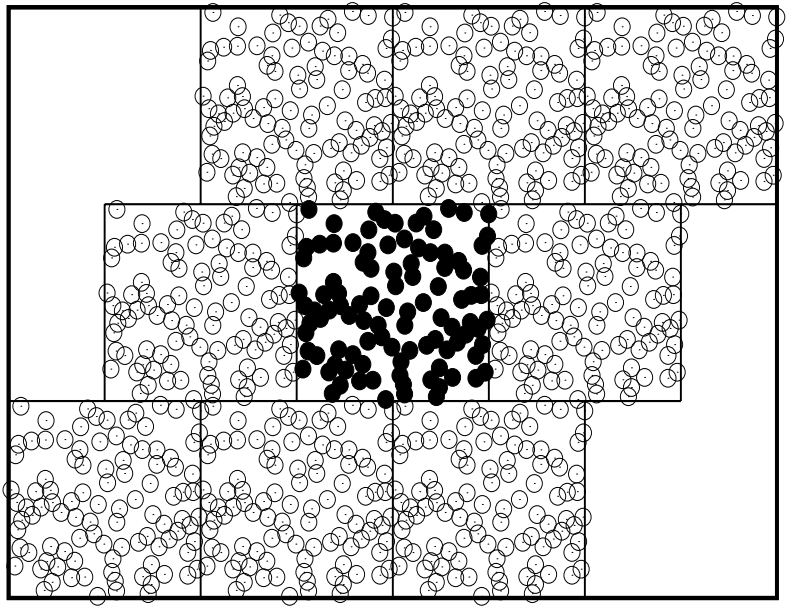}
\caption{
Snapshot from a molecular dynamics simulation of steady periodic shear in two
space dimensions.  The three periodic images of the central $L \times L$ cell shown in the top
row move to the right with velocity $+\dot \epsilon L$ relative to the central 100-particle
cell.  The three images in the bottom row move to the left with velocity $-\dot \epsilon L$ .
The regular checkerboard arrangement of the cells recurs at integral multiples of the
shear period, $\tau = (1/\dot \epsilon) $ .
}
\end{figure}

Steady homogeneous shear flows with periodic boundaries would seem to be the simplest conceivable
nonequilibrium steady states.  Figure 1 shows a particle snapshot taken from a typical molecular
dynamics simulation.
In that Figure the central cell, containing $N = 100$ two-dimensional particles, is repeated
periodically in space.  Periodic images to the right and to the left are generated by adding or
subtracting multiples of the cell side length $L$ to all the particle coordinate values from the central
cell.  100-particle images just  above the central cell move to the right with an additional velocity
$+\dot \epsilon L$ relative to the central row of cells.  Images just below that row likewise move
to the left with an additional velocity $-\dot \epsilon L$.  These periodic boundaries impose a
shear strain rate and a corresponding stress on {\it all} cells.  The {\it strain rate} $\dot \epsilon$
defined and imposed by these periodic boundary conditions is the overall macroscopic velocity
gradient :
$$
\dot \epsilon \equiv \langle \ du_x/dy \ \rangle \ .
$$

Such early 100-particle studies from the 1970s were, by the 1990s, augmented by larger-scale
simulations with as many as $514^2 = 264,196$ particles\cite{b1,b2,b3,b4,b5}.  These
{\it steady-state} flows, in which the heat generated by the shearing motion is extracted by
``thermostat'' (or ergostat) forces, established that the measured shear viscosity, at fixed
strain rate, is insensitive to system size.

Consider a prototypical dense soft-disk fluid, with $N=L^2$ and a short-ranged repulsive pair
interaction $\phi(r<1) = 100(1-r^2)^4$.  For  $N$ greater than 64 the shear viscosity has only
a small ``size dependence''.  For a fixed strain rate the finite-$N$ viscosity
$\eta(\dot \epsilon,N)$ approaches the large-system limit from below.  The $N$-dependent
deviations from the limiting viscosity, $\eta(\dot \epsilon,N\rightarrow \infty)$ , are of order
$\sqrt{(1/N)}$ . The deviation is about two percent for $N = 64$ and is negligible, relative to
statistical fluctuations, for systems with $N$ of order $10,000$ \cite{b3,b4}.

One might well expect, by analogy with macroscopic three-dimensional hydrodynamic flows\cite{b12},
that a {\it turbulent} instability would appear at a Reynolds' Number of the order of a thousand,
$R \equiv \dot \epsilon L^2/(\eta /\rho) \simeq 1000$ .   In fact, detailed investigations, for
thermostated two-dimensional flows with Reynolds' Numbers as large as 50,000 , showed {\it no
instability whatever}.  These molecular dynamics simulations showed instead that
steady two-dimensional dense-fluid shear flows are stable.  These results were announced
in 1995 in two papers by author Bill and Harald Posch, ``Large-System Hydrodynamic Limit''
and ``Shear Viscosity {\it via} Global Control of Spatiotemporal Chaos in Two-Dimensional
Isoenergetic Dense Fluids''.

Soon afterward two papers suggesting a long-wavelength exponential
{\it instability} of {\it all} such flows appeared in the Condensed Matter arXiv: ``Stability
of Uniform Shear Flow'' and ``Long Wavelength Instability for Uniform Shear Flow'', by Jim Dufty,
Mirim Lee, Jim Lutsko, Jos\'e Montanero, and Andr\'es Santos\cite{b6,b7}.  These papers considered
linear perturbations of the density, velocity, and energy fields from the standpoint of the
Navier-Stokes-Fourier equations, specialized to small strain rates $\dot \epsilon$ and to long
wavelengths (or small $k$ vector) normal to the flow direction $(\lambda = 2\pi /k)$ .  Their
analysis showed that for {\it any} fixed strain rate exponential growth invariably results for
perturbations of sufficiently long wavelength.  

In early 2012 we chanced upon these contradictory results [ {\it stable} molecular dynamics and
{\it unstable} linear perturbation theory, both for the {\it same} problem ] .  To address this
puzzle here we first investigated solutions of gas-phase Navier-Stokes-Fourier shear flows with moving
periodic images of a square Eulerian continuum grid, using the same periodic-shear boundary
conditions illustrated in Figure 1 .

\subsection{Continuum Algorithm for Steady-State Simple Shear}

The finite-difference numerical work\cite{b5,b8,b9} is most naturally carried out for grid space
and time differences $\Delta x$ and $\Delta t$ chosen near the Courant condition limit,
$\Delta t < \Delta x/c$ , where $c$ is the fluid's sound speed .  The equations to be solved are
minor modifications of the usual conservation laws for mass, momentum, and energy :
$$
\dot \rho = -\rho\nabla \cdot u \ ; \ \rho \dot u = -\nabla \cdot P \ ; \ \rho \dot e =
-P:\nabla u - \nabla \cdot Q \ .
$$
The boundaries, where the square middle-row cells of Figure 1 contact the
upper and lower rows of cells, need to be made consistent with steady shear. 
Here $\rho$ is the mass density, $u = (u_x,u_y)$ is the velocity, $P$ is the pressure tensor,
$e$ the energy per unit mass, and $Q$ the heat flux vector.  The superior dots indicate ``comoving"
``Lagrangian" time derivatives following the motion.  An efficient algorithm solving this problem
can be based on those developed for the Rayleigh-B\'enard continuum convection problems described
in References 5, 8, and 9.

In the continuum periodic-shear algorithm we added a ``global" ergostat to the energy equation to
constrain the total internal energy.  We also added two global ``thermostats'' to the continuum
equations of motion.  These thermostats constrained the summed-up squares of the $x$ and $y$ 
velocity fluctuations to small constant values, relative to the mean simple-shear motion
$u_x(y) = \dot \epsilon y$ . We also included a short-ranged ``artificial viscosity'' to avoid an
``even-odd'' instability.  The one-dimensional analog of this instability would lead to alternating
signs for velocities evaluated at neighboring nodes of a periodic computation mesh with $N$ zones
defined by $N$ nodes.  At the end of each timestep the $i$th {\it nodal} velocity is replaced by a
weighted average $(0 < \alpha < 1)$ including the {\it zone} velocities of its neighboring zones :
$$
u^n \rightarrow \alpha u_i^n + (1 - \alpha)(u_i^z + u^z_{i-1})/2 \ .
$$

The two-dimensional periodic shear problem is implemented with densities evaluated in $N$ computational
square $\Delta x \times \Delta x$ zones with velocities and energies evaluated at the nodes defining
the four corners of the zones.
Fourth-order Runge-Kutta integration of the $4 \times N = 4L^2$
ordinary finite-difference equations results in a natural strain rate ,
$$
\dot \epsilon = 2\Delta x/(L\Delta t) \ .
$$
The factor 2 reflects the two increments of time occuring in a single Runge-Kutta time step.
This strain rate is of the same order as the $k$ vector
describing the longest-wavelength perturbation :
$$
k \equiv (2\pi/L) \simeq \pi \dot \epsilon (\Delta t/\Delta x) \simeq \pi \dot \epsilon/c \ ,
$$
and is also fairly close to the critical wavelength of the instability analyses.

\subsection{Typical Numerical Results for Continuum Shear of a Gas}

A typical $512 \times 512$ zone simulation, with
$$
[ \ \Delta x = 1 \rightarrow L = 512\ ; \ \Delta t = 0.25 \rightarrow
\dot \epsilon = 2\Delta x/(L\Delta t) = (1/64) \ ] \ ;
$$
$$
[ \ (PV/NkT) = (E/NkT) = 1 \rightarrow c = \sqrt{2} \ ;
\ \nu = (\eta/\rho) = \rho = \eta = \kappa = 1 \ ; \ \eta_V = 0 \ ] \ , 
$$
corresponds to a shear flow with a Reynolds' number
$$
R = L^2\dot \epsilon/\nu = 512 \times (512/64)/\nu  = 4096 \ ,
$$
and is perfectly stable, for a run time of  $256 \times 50$ steps of 0.25 each, corresponding
to a total shear of 50 .

\begin{figure}
\includegraphics[width=4in,height=5in,angle= -90]{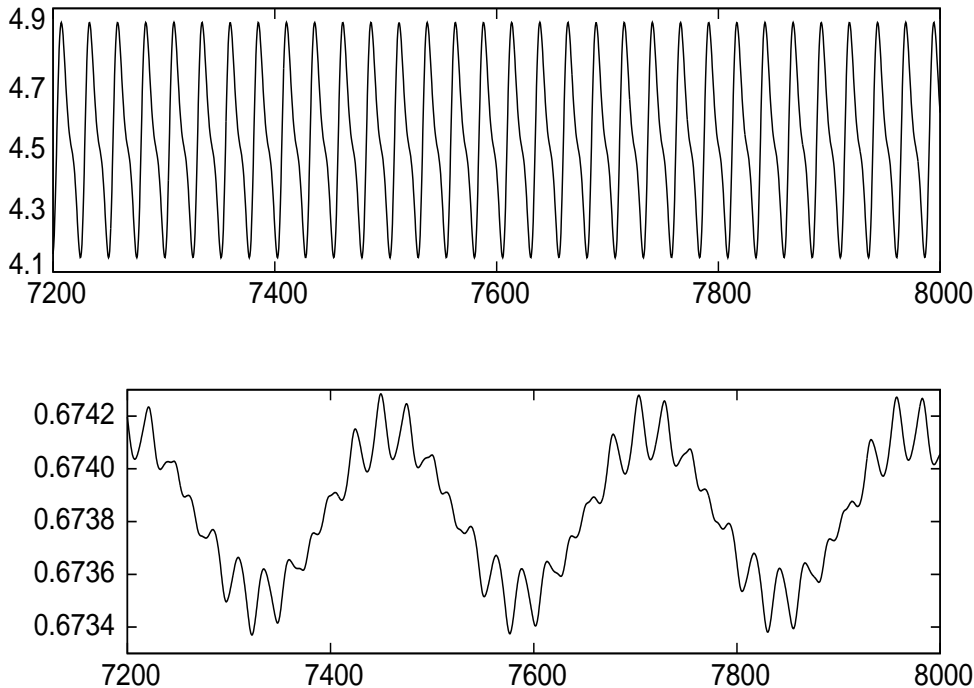}
\caption{
Time history of the horizontal kinetic energy, which includes the single-cell contribution
of the shear to the kinetic energy (below) and the much smaller vertical
kinetic energy (above) in a square $32 \times 32$ Eulerian unit cell.  The
vertical kinetic energy scale has been multiplied by $10^8$ .  The strain rate
is $\dot \epsilon = (2\Delta x/\Delta t)/L = (2/(1/2))/32$ .  A portion of the
time history corresponding to strains from 900 to 1000 is shown.  The transport
coefficients, density, and internal energy, $\{ \ \eta, \kappa, \rho, e \ \}$, are all
equal to unity.  The Reynolds Number is $R = 4 \times 32 = 128$ .
}
\end{figure}

We carried out a variety of simulations.  Time-dependent data from one of them are shown in
Figures 2 and 3 .
The kinetic energy history, after an irregular transient, is typical, with nearly periodic
oscillations of the horizontal and vertical velocity deviations from simple shear flow [ these
deviations are rescaled
at the conclusion of each timestep ] .  The deviations are analogous to thermal fluctuations.
Figure 3 shows these same fluctuations as arrows.  These patterns are not stationary, but vary with
time.  Because the time periodicity of the shear, with period $\tau = (1/\dot \epsilon)$, is
built into the boundary conditions an inhomogeneous perturbation flow is to be expected.

We explored systems with $L \times L$ computational zones, with $L$ = 8, 16, 32, 64, 128, 256,
and 512 to strains of at least 100.  With $\Delta x = 1$ and $\Delta t = (1/2)$ , so that the
Reynolds number is $R=4L$ ,
all members of this series were stable, giving results similar to those in Figures 2
and 3 .  Somewhat larger or smaller values of $\Delta t$ typically led to instability, making
it difficult to validate the stability work of References 6 and 7 .

\begin{figure}
\includegraphics[width=4in,height=4in,angle= -90]{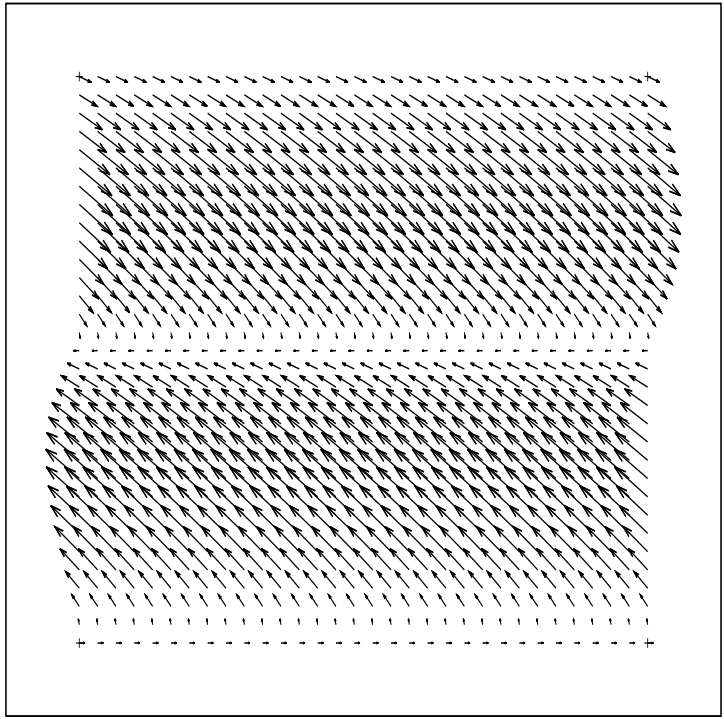}
\caption{
The velocity, relative to the mean flow, $\dot x = \dot \epsilon y$ for the
$32 \times 32$ grid described in Figure 2.  Although this velocity perturbation
is not stationary -- it changes with time -- this view is typical.
}
\end{figure}

These exploratory calculations could be extended in {\it many} arbitrary ways.  We chose
instead to follow and extend the linear perturbation-theory hard-sphere work of Dufty
{\it et alii}\cite{b6,b7} directly to a numerical model based on our earlier soft-disk
shear-flow model.  Our results, described in what follows, suggest that the linear perturbation
analysis yields flawed predictions, contradicted by the well-established stability of thermostated
simple shear in two space dimensions\cite{b2,b3,b4}.  Despite our efforts to extend the
perturbation analyses, described below, a convincing theoretical
demonstration of this finding remains a challenging research goal.
It seems likely that the flaw in the perturbation treatment is an inadequate treatment of steady-state
thermostating.  Our checks of the other aspects of the perturbation theory have so far revealed no errors. 

\section{Constitutive Model for Soft Disks}

The short-ranged repulsive potential $\phi(r<1) = 100(1 - r^2)^4$ has been carefully investigated
near the reference state, $[ \ \rho = 1 \ ; \ e = 1 \ ]$ , where $e$ is the energy per particle and each
particle has unit mass\cite{b3}.  The equilibrium temperature and pressure at that state are
approximately $T$ = 0.69 and $P$ = 3.90 .  For our numerical stability studies we adopt the
equation of state from Reference 3 :
$$
P/\rho = 5 + 8(\rho -1) + 2.5(e - 1.443) + 9(\rho - 1)^2 + 2(\rho - 1)(e - 1.443) \ ;
$$
$$
T = 1 - (\rho - 1) + 0.7(e - 1.443) - 0.8(\rho - 1)^2 - 0.5(\rho - 1)(e - 1.443) \ ;
$$
$$
e - 1.443 = 1.5(\rho - 1) + 1.5(T - 1) + 2.4(\rho - 1)^2 + 1.2(\rho - 1)(T - 1) \ .
$$
We estimate the transport coefficients (and evaluate their density and temperature derivatives
numerically) $\eta, \ \eta_V, \ \kappa$ from Enskog's theory\cite{b10,b11}.

Enskog suggested that transport properties for soft potentials be estimated from those of a corresponding
hard-potential model.  The equivalence is based on setting the soft model's ``thermal pressure'',
$T(\partial P/\partial T)_V$, equal to the hard-potential pressure.  In our two-dimensional implementation
the hard-disk model corresponding to the soft-disk potential $\phi = 100(1-r^2)^4$ follows from the
relation :
$$
(V/Nk)(\partial P/\partial T)_{\rm soft} \equiv (PV/NkT)_{\rm hard} \ .
$$
For the hard-disk compressibility factor we use the ten-term virial series from Clisby and McCoy's
work\cite{b10} :
$$
[ \ (\partial P/\partial T)/\rho \ ]_{\rm soft} \equiv (P/\rho T)_{\rm hard} = 1 + (b\rho)\chi =
1 + b\rho + 0.782(b\rho)^2 + 0.532(b\rho)^3 +
$$
$$
 0.334(b\rho)^4 + 0.199(b\rho)^5 + 0.115(b\rho)^6 +
0.065(b\rho)^7 + 0.036(b\rho)^8 + 0.020(b\rho)^9
+ \dots \ .
$$
Here $\chi$ is the ratio of the actual hard-disk collision rate from the virial series to the
low-density kinetic-theory collision rate calculated with the two-term series.  The computed value
of the hard-disk second virial coefficient $b = \pi \sigma^2/2$ gives the hard-disk diameter
$\sigma = \sqrt{2b/\pi}$ needed for the hard-disk transport coefficients.

David Gass kindly worked out formul\ae$\ $ for the Navier-Stokes-Fourier transport
coefficients for hard disks in 1970\cite{b11} :
$$
\eta = \eta_0 b \rho [ \ (b \rho \chi)^{-1} + 1.0 + 0.8729(b \rho \chi) \ ] \ ;
\ \eta_V = 1.246 \eta_0 b \rho (b \rho \chi) \ ; 
\ \eta_0 = \sqrt{ mkT/\pi } /2\sigma \  ;
$$
$$
\kappa = \kappa_0 b \rho [ \ (b \rho \chi)^{-1} + 1.5 + 0.8718(b \rho \chi ) \ ] \ ;
\ \kappa_0 = 2\sqrt{k^3T/m\pi}/\sigma \  .
$$
These expressions provide estimates for the soft-disk transport coefficients needed to apply
the perturbation theory described in the following Section.  

\section{Dufty-Lee-Lutsko-Montanero-Santos' Linear Theory}

Consider the reference steady shear flow state for soft disks at unit density and energy :
$$
\{ \ \rho = 1 \ , \ u_x  = \dot \epsilon y \ , \ u_y = 0 \ , \ e = 1 \ \} \ .
$$
There are extensive molecular dynamics simulation data available for strain rates $\dot \epsilon$ of
0.10, 0.25, and 0.50 .  We apply linear perturbation theory to this problem for a
perturbation with components
$$
\{ \ \delta \rho \times e^{ iky + i \omega t} \ , \ \delta u_x \times e^{i ky + i \omega t} \ , \
\delta u_y \times e^{i ky + i \omega t} \ , \ \delta e \times e^{i ky + i \omega t} \ \} \ ,
$$
where the unperturbed state is steady shear at a mass density and internal energy per particle of
unity :
$$
\{ \ \delta \rho = \rho - 1 \ , \ \delta u_x = u_x - \dot \epsilon y \ ,
\ \delta u_y = u_y \ , \ \delta e = e - 1 \ \} \ .
$$
In addition to the conservation laws we use the Navier-Stokes-Fourier constitutive relations 
$$
P = [ \ P_{\rm eq} - \lambda \nabla \cdot u \ ] I - \eta [ \ \nabla u + \nabla u^t \ ] \ ; \ Q = -\kappa \nabla T \  ,
$$
with the Gass-Enskog transport coefficients.  As usual $\eta $ is the shear viscosity. Here $\lambda $
is the difference between the bulk and shear viscosities , $\lambda = \eta_V - \eta $ , and $\kappa $
is the heat conductivity.  To a good approximation $\lambda \simeq 0$, so that the bulk and shear
viscosities are roughly equal, and about four times less than the heat conductivity \ :
$$
\eta \simeq 1 \simeq \eta_V = \lambda + \eta  \ ; \ \kappa \simeq 4\eta \ .
$$

\subsection{Matrix Solution of the Linear Equations}

If we represent the hydrodynamic state perturbations by the four-dimensional vector $\delta$ then the linearized hydrodynamic equations take the form of a $4 \times 4$ matrix equation,
$\dot \delta = M \cdot \delta$.  Let us begin by estimating the values of the 16 matrix elements.
If we (1) ignore the {\it nonlinear} convective contributions to the equations of motion and (2) assume a
solution of the form $\exp[ \ iky + i\omega t \ ]$ we get a set of four {\it linear} equations for the
evolution, in space and time, of perturbations in mass, momentum, and energy : \\

\noindent
$
{\rm [ \ mass \ conservation \ ] \ : \ } \dot \rho = - \rho \nabla \cdot u \longrightarrow
$
$$
(\partial \delta \rho/\partial t) \equiv  i \omega \delta \rho =
-\rho (\partial \delta u_y/\partial y) \equiv - ik\rho\delta u_y \ ;
$$
\noindent
{\rm [ } $x$ {\rm momentum conservation ] : }   $\dot u_x = -\dot \epsilon \delta u_y - (1/\rho)(\partial P_{xy}/\partial y) \longrightarrow
$
$$
i\omega \delta u_x = - \dot \epsilon \delta u_y - (1/\rho)(\partial /\partial y)[ \ -\eta (\partial /\partial y)(\dot \epsilon y + \delta u_x) \ ] \simeq
$$
$$
-\dot \epsilon \delta u_y +
ik(\dot \epsilon/\rho)[ \ (\partial \eta/\partial \rho)_e\delta \rho +
(\partial \eta/\partial e)_\rho \delta e \ ] - (\eta/\rho)k^2\delta u_x  \ .
$$
\noindent
{\rm [ } $y$ {\rm momentum \ conservation \ ] : }  $\dot u_y = - (1/\rho)(\partial P_{yy}/\partial y) \longrightarrow
$
$$
i\omega \delta u_y = - (1/\rho)(\partial /\partial y)[ \ P_{eq} - (\eta + \eta_V)(\partial u_y/\partial y) \ ] \simeq
$$
$$
-(ik/\rho)[ \ (\partial P_{eq}/\partial \rho)_e\delta \rho +
(\partial P_{eq}/\partial e)_\rho\delta e \ ] - (\eta + \eta_V)k^2\delta u_y \ .
$$
$
{\rm [ \ energy \ conservation \ ] \ : \ } \dot e = (-1/\rho)[ \ P:\nabla u + \nabla  \cdot Q \ ] =
$
$$
 -(1/\rho)[ \ P_{xy}(\partial u_x/\partial y) + P_{yy}(\partial u_y/\partial y) +
(\partial Q_y/\partial y) \ ] \longrightarrow
$$
$$
i\omega \delta e \simeq
\dot \epsilon^2[ \ (\partial \nu/\partial \rho)_e\delta \rho +
(\partial \nu/\partial e)_\rho\delta e \ ] +
2ik\nu \dot \epsilon\delta u_x
$$
$$
 -ik(P_{eq}/\rho)\delta u_y - k^2(\kappa/\rho)\delta T \ .
$$
Note that the kinematic viscosity, $\nu \equiv (\eta/\rho)$ has been introduced in the energy conservation
expression.  For computation of the matrix elements, and their eigenvalues, it is necessary to express $\nabla ^2T$ in terms of $\delta \rho$ and $\delta e$ :
$$
\delta T = -0.78\delta \rho + 0.7\delta e \longrightarrow \nabla^2T \simeq k^2[ \ 0.78 \delta \rho
- 0.7 \delta e \ ] \ .
$$
A numerical evaluation of the eigenvalues for a dense grid of $k$-vectors and strain rates gives
two unstable ( positive real part, leading to exponential growth ) eigenvalues in the shaded
lower-triangular region shown in Figure 3 ( which closely resembles the behavior found by Dufty {\it et
alii} ).  Notice that the stable large-system molecular dynamics data from References 3 and 4, shown also in
the Figure, lie squarely inside the ``unstable region".

The contradiction between perturbation theory and molecular dynamics is robust.  We have
investigated several modifications of the basic theory and found qualitatively similar results.
Replacing the ``Sllod'' matrix element with its ``Doll's-Tensor'' analog,
$$
\{ \ M_{u_x,u_y} = -\dot \epsilon \ ; \ M_{u_y,u_x} = 0 \ \} \longleftrightarrow 
\{ \ M_{u_x,u_y} = 0 \ ; \ M_{u_y,u_x} = -\dot \epsilon \ \} \ ,
$$
leaves the qualitative picture unchanged.  Reducing the matrix to $3 \times 3$ by ignoring the
possibility of energy fluctuations still yields an unrealistic triangular ``unstable region''.
We have no explanation for the failure of perturbation theory to deal with steady simple shear.

\begin{figure}
\includegraphics[width=2.5in,angle= -90]{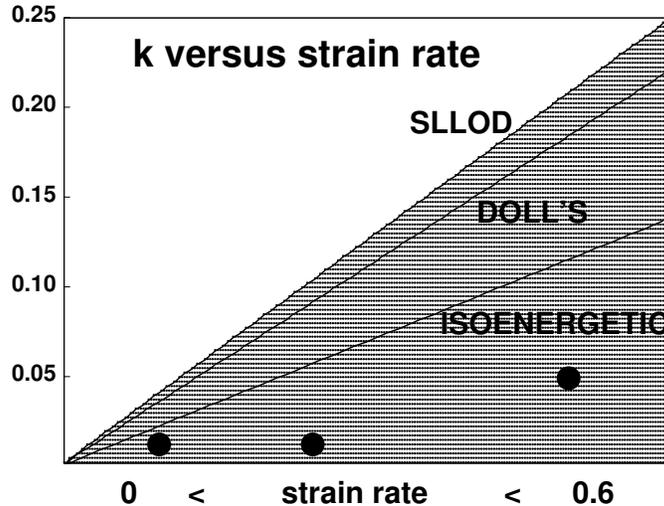}
\caption{
Portion of $k$-vector {\it versus} strain rate space for two-dimensional soft disks.
Flows in a lower triangular region [ shading shows that region for the Sllod Matrix ] are
{\it unstable}, leading to longtime exponential growth
of perturbations, according to the theory outlined in References 6 and 7.  Stable
atomistic simulations for soft disks have been carried out for the conditions corresponding
to the three filled circles (as well as for {\it many} other situations).  The two approaches
to stability give contradictory results.  The straightforward nature of the molecular dynamics
argues that the perturbation-theory results are invalid.
}
\end{figure}

\section{Conclusion}

Although the evidence from molecular dynamics supports the large-system stability of thermostated 
steady shear, linear stability analysis, with a fixed global thermostat, predicts {\it instability}.
Where might the explanation lie?  We don't know.  Explanations might involve either of
two flow features treated poorly by the theory: [1] The ``steady-state" shear flow is not
really stationary.  See again Figure 2.  At times which are integral multiples of the inverse
shear rate,
$t = n(1/\dot \epsilon)$, the periodic arrangement of cells is a perfect rectangular
checkerboard; [2] Evidently the molecular dynamics thermostats provide additional
stability to the flow.

A variety of molecular dynamics algorithms, with thermostats or
with ergostats and with Doll's or Sllod equations of motion, all give very similar results,
all of them identical for the {\it linear} response proportional to $\dot \epsilon$ .  It may be
that some more faithful representation of the thermostat forces in the perturbation theory
could lead to results consistent with the molecular dynamics.  We would be most grateful
for an explanation of the apparent contradiction between the atomistic and continuum
approaches to the stability of steady two-dimensional ``simple shear".

It seems to be an accepted result (though the exact assumptions required are unclear) that 
{\it linear} stability theory predicts stability for simple shear at {\it all} Reynolds
numbers.  See, for instance, Reference 12.  On the other hand the necessary {\it inhomogeneity}
due to the boundary conditions, and the need for thermostat and/or ergostat heat sinks, provide
opportunities for a wide variety of much more complex models for this ``simple'' shear problem.

\section{Acknowledgment}

Bill thanks Andrew Gabriel De Rocco for arranging a meeting with Peter Debye {\it circa} 1960.
Debye's advice, to choose research topics from among existing puzzles in the literature, has
been an enduring guide.  We are looking forward to discussing the present work at the 40th
Summer School, ``Advanced Problems in Mechanics'' at Saint Petersburg in Summer 2012 .

\section{Addendum of 24 April 2012}

We wish to thank Vitaly Kuzkin for emphasizing the relevance of Denis Evans and Gary Morriss' paper, ``Nonequilibrium Molecular Dynamics Simulation of Couette Flow in Two-Dimensional Fluids", Physical Review Letters {\bf 51}, 1776-1779 (1983).  Evans and Morriss studied the shear viscosity of 896 and 3584 soft-disk particles at several strain rates and densities.  Just as in our work in References 1-5 they observed only a small number dependence, and a nearly rate-independent viscosity rather than the divergence of viscosity predicted by the mode-coupling models.  Their viscosity data look closely Newtonian, so long as the strain rate is not too large.  In 1983 this was a surprise as the mode-coupling models of the 1970s current then predicted the {\it divergence} of viscosity in two-dimensional fluids.  Influenced by those models, Evans and Morriss explained the observed unexpected stability of their own results as a ``screening instability" due to transverse velocity fluctuations.  By removing the three longest-wavelength fluctuations Evans and Morriss recovered a logarithmic rate-dependence, $\eta \simeq \ln (1/\dot \epsilon)$.  They commented that larger systems would require the removal of even more Fourier components of the transverse velocity fluctuations in order to squelch their observed Newtonian behavior.


\begin{thebibliography}{99}

\newpage

\bibitem{b1}  For an early review see Wm. G. Hoover and W. T. Ashurst, ``Nonequilibrium
Molecular Dynamics'' in {\it Theoretical Chemistry, Advances and Perspectives} {\bf 1}, 1-51 (1975).

\bibitem{b2}  For a more recent survey see Wm. G. Hoover, C. G. Hoover, and J. Petravic,
``Simulation of Two- and Three-Dimensional Dense-Fluid Shear Flows {\it via} Nonequilibrium
Molecular Dynamics: Comparison of Time-and-Space-Averaged Stresses from Homogeneous Doll's
and Sllod Shear Algorithms with Those from Boundary-Driven Shear'', arXiv 0805.1490, Physical Review E
{\bf 78}, 046701 (2008).

\bibitem{b3}  Wm. G. Hoover and H. A. Posch, ``Shear Viscosity {\it via} Global Control of
Spatiotemporal Chaos in Two-Dimensional Isoenergetic Dense Fluids", Physical Review E
{\bf 51}, 273-279 (1995) .

\bibitem{b4}  Wm. G. Hoover and H. A. Posch, ``Large-System Hydrodynamic Limit''
Molecular Physics Reports {\bf 10}, 70-85 (1995).

\bibitem{b5} Wm. G. Hoover and C. G. Hoover, {\it Time Reversibility, Computer Simulation,
Algorithms, Chaos} (World Scientific, Singapore, 2012).

\bibitem{b6} M. Lee, J. W. Dufty, J. M. Montanero, A. Santos, and J. F. Lutsko,
``Long Wavelength Instability for Uniform Shear Flow'', Condensed Matter arXiv 9604187,
Physical Review Letters {\bf 76}, 2702-2705 (1996).

\bibitem{b7} J. M. Montanero, A. Santos, M. Lee, J. W. Dufty, and J. F. Lutsko,
``Stability of Uniform Shear Flow'', Condensed Matter arXiv 9705168,
Physical Review E {\bf 57}, 546-556 (1998).   

\bibitem{b8} M. Mareschal, M. Mansour, A. Puhl, and E. Kestemont, ``Molecular Dynamics
{\it versus} Hydrodynamics in a Two-Dimensional Rayleigh-B\'enard System'', Physical
Review Letters {\bf 61}, 2550-2553 (1988).  

\bibitem{b9} A. Puhl, M. Mansour, and M. Mareschal, ``Quantitative Comparison of Molecular
Dynamics with Hydrodynamics in Rayleigh-B\'enard Convection'', Physical Review A {\bf 40},
1999-2012 (1989).

\bibitem{b10} N. Clisby and B. M. McCoy, ``Ninth and Tenth Order Virial Coefficients for Hard
Spheres in $D$ Dimensions", Condensed Matter arXiv 0503525, Journal of Statistical  Physics
{\bf 122}, 15-57 (2006).

\bibitem{b11} D. M. Gass, ``Enskog Theory for a Rigid Disk Fluid", Journal of Chemical
Physics {\bf 54}, 1898-1902 (1971).

\bibitem{b12} D. Viswanath, ``The Dynamics of Transition to Turbulence in Plane Couette
  Flow'', Fluid Dynamics arXiv 0701337, in {\it Mathematics and Computation: a Contemporary
View}, The 2006 Abel Symposium ( Springer-Verlag, Berlin, 2008).
 
\end{thebibliography}
\end{document}